\documentclass[aos]{imsart}

\RequirePackage{amsthm,amsmath,amsfonts,amssymb}
\RequirePackage[numbers]{natbib}
\usepackage{algorithm}
\usepackage{bm,bbm}
\usepackage{graphicx}
\usepackage{xcolor}
\RequirePackage[colorlinks,citecolor=blue,urlcolor=blue]{hyperref}

\startlocaldefs
\def\P{\mathbb{P}}
 \def\E{\mathbb{E}}
 \def\R{\mathbb{R}}
 \def\eps{\epsilon}
\def\bx{\bm{x}}

\def\calX{\mathcal{X}}

\def\calD{\mathcal{D}}

\def\sig{\sigma}
\def\btheta{\bm{\theta}}

\def\eps{\epsilon}

\def\ptr{\mathrm{ptr}}

\def\fsp{\textup{fsp}}
\DeclareMathOperator*{\argmin}{arg\,min}

\newtheorem{corollary}{Corollary}

\newtheorem{lemma}{Lemma}
\newtheorem{definition}{Definition}
\newtheorem{condition}{Condition}
\newtheorem{theorem}{Theorem}
\newtheorem{example}{Example}
\newtheorem{remark}{Remark}

\endlocaldefs

\begin{document}

\begin{frontmatter}
\title{Personalizing black-box models for nonparametric regression with minimax optimality}

\runtitle{Few-shot personalization}

\begin{aug}
\author[A]{\fnms{Sai}~\snm{Li}\ead[label=e1]{saili@tsinghua.edu.cn}},
\and
\author[B]{\fnms{Linjun}~\snm{Zhang}\ead[label=e2]{linjun.zhang@rutgers.edu}}
\address[A]{Department of Statistics and Data Science,Tsinghua University\printead[presep={,\ }]{e1}}

\address[B]{Department of Statistics, Rutgers University\printead[presep={,\ }]{e2}}
\end{aug}

\begin{abstract}
Recent advances in large-scale models, including deep neural networks and large language models, have substantially improved performance across a wide range of learning tasks. The widespread availability of such pre-trained models creates new opportunities for data-efficient statistical learning, provided they can be effectively integrated into downstream tasks. Motivated by this setting, we study few-shot personalization, where a pre-trained black-box model is adapted to a target domain using a limited number of samples. We develop a theoretical framework for few-shot personalization in nonparametric regression and propose algorithms that can incorporate a black-box pre-trained model into the regression procedure. We establish the minimax optimal rate for the personalization problem and show that the proposed method attains this rate. Our results clarify the statistical benefits of leveraging pre-trained models under sample scarcity and provide robustness guarantees when the pre-trained model is not informative. We illustrate the finite-sample performance of the methods through simulations and an application to the California housing dataset with several pre-trained models.
\end{abstract}

\begin{keyword}[class=MSC]
\kwd[Primary ]{62G05}
\kwd{68T05}
\kwd[; secondary ]{62C20}
\end{keyword}

\begin{keyword}
\kwd{personalization}
\kwd{nonparametric regression}
\kwd{minimax optimality}
\end{keyword}

\end{frontmatter}
\maketitle
\section{Introduction}

Recently, large-scale black-box models have led to dramatic improvements in predictive performance across various tasks. Powerful pre-trained black-box models create new opportunities for data-efficient learning, provided that these models can be effectively and rigorously integrated into downstream applications. This, in turn, motivates the development of principled statistical frameworks for incorporating pre-trained models into learning procedures.

A central challenge in leveraging pre-trained models across heterogeneous settings is personalization: adapting a large-scale model trained on one population or task to a new one using limited labeled data. \emph{Few-shot personalization} aims to adapt a black-box pre-trained model to new users or tasks using only a small number of target samples \citep{kim2024few}. As a motivating example, NeuralCVD \citep{steinfeldt2022neural} is a neural-network-based model for predicting cardiovascular risk trajectories trained on UK Biobank data. 
When applied to a different population without adaptation, its performance may degrade substantially. However, retraining such models from scratch is often infeasible due to data scarcity or computational constraints. In these settings, adapting an existing pre-trained model using limited target observations offers a natural and efficient alternative.

Formally, let $y\in\R$ denote the response variable and $\bx\in\mathcal{X}\subseteq \R^d$ denote the covariates. Let $f^{(\mathrm{ptr})}:\mathcal{X}\to\mathbb{R}$ be a pre-trained model learned from an external source. Our goal is to construct a personalized predictor for a target distribution $P^*(\bx,y)$ by integrating $f^{(\ptr)}$ with a small number of samples drawn from the target population, without accessing the internal mechanism of $f^{(\ptr)}$. When the pre-trained model carries informative signal for the target task, it is desirable to borrow strength from $f^{(\ptr)}$ and adjust for the discrepancies between the source and target distributions. Our framework allows the pre-trained model to be a black box and takes its predictions as input. We require no access to model parameters, architecture, or any prior knowledge of its relevance to the target problem. 
This setting reflects practical scenarios where users aim to enhance small data analysis using proprietary models (e.g., GPT-4 and Gemini) whose internal workings are inaccessible. Even with open-source models (e.g., Llama), direct fine-tuning may remain computationally prohibitive. Personalization thus offers a computationally efficient alternative for integrating large-scale black-box models.

We study personalization in the nonparametric regression setting. Given a covariate vector $\bx\in\mathcal{X}$, the response in the target population follows
\begin{equation}
\label{m1}
y = f^*(\bx) + \varepsilon, \qquad \mathbb{E}[\varepsilon \mid \bx] = 0,
\end{equation}
where $\varepsilon$ denotes a noise variable with conditional variance $\mathrm{Var}(\varepsilon \mid \bx) = \sigma^2(\bx)$, allowing for heteroskedasticity.
The unknown regression function $f^*: \mathcal{X} \to \mathbb{R}$ is the target of interest. In general, $f^*(\cdot)$ need not coincide with the pre-trained model $f^{({\ptr})}(\cdot)$ and may differ substantially from it. As a result, adaptation using target data is essential to calibrate $f^{(\ptr)}$ to the target population. Since collecting labeled target samples is often costly, we assume a limited sampling budget of size $n$ and propose a sample collection scheme specifically designed to facilitate personalization.

In summary, the personalization procedure takes as input a pre-trained model $f^{(\ptr)}$, a target covariate domain $\mathcal{X}$, and a sampling budget $n$. Based on $n$ labeled observations generated from \eqref{m1}, we develop algorithms to construct a personalized estimator of $f^*(\cdot)$. Extensions of this framework to other settings are discussed in later sections.

\subsection{Connections to related works}

Few-shot personalization has gained importance in large language models \cite{kim2024few} and federated learning \cite{zhao2022personalized}, where the common goal is to adapt a pre-trained model to diverse users or tasks. However, the statistical understandings are limited. The statistical optimal procedures are largely unknown and the minimax optimal rates have not been investigated.
We first review some related topics highlighting the unique challenges of few-shot personalization.

To integrate external information, transfer learning and domain generalization are popular schemes which leverage source data to boost the learning performance of the target data. Many recent works have studied transfer learning approaches for nonparametric regression \cite{cai2021transfer,reeve2021adaptive}, high-dimensional parametric models \cite{li2022transfer,tian2023transfer,li2024estimation} among many others. Domain generalization is another out-of-distribution prediction paradigm, where the focus is to train a prediction rule from multiple source domains that enable it to perform well on unseen target domains. State-of-the-art methods study invariant prediction rules \cite{peters2016causal} in causal and machine learning models. 
Under causal structural models, \cite{peters2016causal,buhlmann2020invariance,rothenhausler2021anchor} develop invariant methods for linear models. Without causal assumptions, \cite{baktashmotlagh2013unsupervised} propose a domain invariant projection method to learn a transformation of covariates which is invariant across domains. \cite{rojas2018invariant} develop the invariant risk minimization framework, aiming to discover invariant representations across multiple training environments while excluding spurious features. \cite{fan2023environment} study an invariant least square approach for domain generalization under infinite sample conditions. Beyond the invariance framework, \cite{li2025multi} study domain generalization when the regression coefficients can be organized as a low-rank tensor. 
To summarize, in transfer learning and domain generalization, users can access individual-level source data to build a desired pre-trained model for best knowledge transfer. In contrast, personalization treats the pre-trained model as a black box with no access to source data, and explicitly incorporates a sample collection phase.

Recently, prediction-powered inference (PPI) framework \cite{angelopoulos2023prediction,angelopoulos2023ppi++} is proposed to combine an arbitrary pre-trained model for statistical inference in the semi-supervised setting. In the PPI framework, $f^{(\ptr)}(\bx)$ can be treated as a surrogate variable for inference of the population-level parameters \cite{ji2025predictions}.
There are two key differences between PPI and personalization. First, PPI targets population means rather than regression functions or individual predictions. Second, PPI considers the semi-supervised setting: integrating the pre-trained models with a few labeled samples and a larger amount of unlabeled samples.
The samples are assumed to be randomly collected from target distribution. In contrast, in the personalization problem, we only leverage a few labeled data. Moreover, the covariate distribution of the labeled data is user-specified which can be different from the target covariate distribution. 
A more closely related work is \cite{zrnic2024active} which studies selecting samples to be labeled in the PPI setting but their focus is still to make inference for the population mean. We propose a different sampling rule tailored for the personalization problem and establish theoretical guarantees for our proposal. 

Recent work on personalizing large language models (LLMs) begin to explore how to tailor model outputs to individual users’ preferences, writing styles, and histories \cite{Zhang2024_PersonalizationLLMs_Survey,liu2025survey}. On the methodological side, \cite{Salemi2024_LaMP} show empirically that retrieval-augmented generation and fine-tuning (or prompting) can significantly improve personalized per-user text classification and generation tasks compared to generic LLMs. More recent algorithmic advances also propose to learn user-specific prompts \cite{KimYang2024_FERMI} or user-specific reward \cite{shenfeld2025language} to steer LLM outputs toward individual preferences using only a small number of user feedbacks.  Despite this growing diversity, these contributions are primarily empirical or algorithmic, lacking formal statistical analysis of adaptation under limited data and minimax optimality guarantees.

 \subsection{Main results}

In this work, we present a statistical framework and efficient algorithms for personalization in non-parametric regression. Our contributions are twofold.
 
First, we propose a personalization procedure that integrates a black-box pre-trained model into a nonparametric regression estimator through local smoothing. This step plays a central role in achieving robustness to model misspecification while maintaining statistical efficiency. In addition, we introduce a sample retrieval scheme designed to improve sampling efficiency in the presence of heteroskedastic noise. Together, these components provide a principled approach for incorporating large-scale pre-trained models into classical nonparametric inference.

On the theoretical side, we characterize the minimax optimal rate for personalized nonparametric regression and show that the proposed estimator attains this rate under mild conditions. Our results characterize how the prediction accuracy depends on the H\"older norm of the target regression function in the minimax sense. 
The analysis reveals how leveraging a pre-trained model can reduce the H\"older complexity of the estimation problem and, in some regimes, improve the effective smoothness order, thereby yielding faster convergence rates compared to target-only estimation. 
Importantly, the proposed method enjoys a \emph{no-harm guarantee}: its performance is provably no worse than that of a nonparametric estimator based solely on target samples under mild conditions, regardless of the pre-trained model's relevance. To our knowledge, this is among the first works to provide a minimax-optimal statistical treatment of personalization with rigorous guarantees.

\subsection{Organization and notation}

The remainder of the paper is organized as follows. 
In Section~\ref{sec-np}, we introduce the proposed few-shot personalization method for nonparametric regression. Section~\ref{sec-theory} presents theoretical analysis and minimax optimal rates. In Section~\ref{sec-xstar}, we extend the framework to settings where additional unlabeled covariates from the target domain are available. Section~\ref{sec-simu} reports simulation studies comparing the proposed method with existing approaches, and Section~\ref{sec-data} applies the method to a real data analysis involving prediction of California housing prices. Proofs and technical details are deferred to the supplement.

For a set $\mathcal{S}\in\R^d$, let $|\mathcal{S}|$ denote its Lebesgue measure. 
Let $a_n=O(b_n)$ and $a_n\lesssim b_n$ denote $|a_n/b_n|\leq c$ for some constant $c$ when $n$ is large enough. Let $a_n=o(b_n)$ and $a_n\ll b_n$ denote $a_n/b_n\rightarrow 0$ as $n\rightarrow\infty$. We use $C,C_0,C_1,\dots,c,c_0,c_1,\dots$ to denote generic constants which can be different in different statements.

\section{Personalized estimation of the nonparametric function} 
\label{sec-np}

This section introduces our proposed method for personalizing a pre-trained model $f^{(\mathrm{ptr})}(\cdot)$ to estimate the target regression function $f^*(\bx)$ over a domain $\mathcal{X} \subseteq \mathbb{R}^d$. We first define the function class under consideration and then present the complete procedure.

\begin{definition}[H\"older class]
\label{def1}
For $\bm\theta=(\theta_1,\theta_2)^{\top}$, the H\"older class $H(\bm\theta)$ on $\mathbbm{T}\subseteq \R^d$ for some finite $\theta_1\geq 0$ and $0<\theta_2\leq 1$, is defined as the set of functions $f: \mathbbm{T}\rightarrow \R$ satisfying, for any $\bx_1,\bx_2\in \mathbbm{T}$,
\[
   |f(\bx_1)-f(\bx_2)|\leq \theta_1\|\bx_1-\bx_2\|_2^{\theta_2}.
\]
If $f\in H(\bm\theta)$, we call $\btheta$ the H\"older class parameters of $f(\cdot)$.
\end{definition}
In Definition \ref{def1}, we focus on H\"older classes with smoothness order $\theta_2 \le 1$, which are standard in nonparametric regression literature \cite{fan1993local, tsybakov2009nonparametric} and transfer learning literature \cite{cai2021transfer}. 
Unlike classical theories that assume the  H\"older norm $\theta_1$ is fixed, we only require $\theta_1$ to be finite and allow it to go to zero as sample size $n\rightarrow\infty$. 
This flexibility enables a more refined characterization of how smoothness parameters influence the convergence rates and plays an important role in our personalization analysis. 

Throughout, we assume that the target regression function satisfies $f^*(\bx)\in H(\btheta^*)$ on $\mathcal{X}$ for some finite $\theta^*_1>0$ and $0<\theta^*_2\leq 1$. For simplicity, we take $\mathcal{X}=[0,1]^d$ and an extensions is studied in Section \ref{sec1-xstar}. 
As for the risk criteria, we define
 the mean integrated squared error for a generic function $f:\mathcal{X}\rightarrow \R$ as
\begin{align}
\label{mise}
  \text{MISE}(f)=\E\left[\int_{\mathcal{X}} \{f(\bx)-f^*(\bx)\}^2d\bx\right],
  \end{align}
  which is a standard metric in the classical nonparametric literature \citep{tsybakov2009nonparametric}.

\subsection{Overview of the main steps}
\label{sec2-1}
We begin by outlining the proposed personalization procedure. The method consists of three main steps.
\begin{itemize}
\item \textbf{Step 1: Sample retrieval.} We first retrieve $n$ covariate points $\bx_i\in \calX$, $i=1,\dots,n$, according to a data collection rule specified in Section~\ref{sec-sample}. For each selected $\bx_i$, we collect the corresponding response $y_i$ generated from the target model \eqref{m1}. 
\item \textbf{Step 2: Smoothed bias correction.} We calibrate the pre-trained model $f^{(\ptr)}(\bx)$  by estimating its bias function $\delta(\bx)=f^*(\bx)-f^{(\ptr)}(\bx).$ Since $\delta(\bx)$ need not be smooth, directly estimating it may be unstable. Instead, we first apply a local smoothing operation, called $\btheta$-local-smoothing, to the pre-trained model, indexed by a tuning parameter $\btheta$. Then we construct a kernel-based estimator of the resulting bias, denoted by $\hat{\delta}_{{\btheta}}(\bx)$. The corresponding personalized estimator is 
$$\hat{f}^{(\fsp)}_{\btheta}(\bx)= f^{(\ptr)}(\bx)+\hat{\delta}_{\btheta}(\bx).$$
\item \textbf{Step 3: Adaptation.} We select the tuning parameter $\hat{\btheta}$ using validation samples and output the final few-shot personalized estimator $$\hat{f}^{(\fsp)}(\bx)= \hat{f}^{(\fsp)}_{\hat{\btheta}}(\bx).$$
\end{itemize}
This algorithm's core idea is to correct the bias of the pre-trained model in a cost-effective and statistically efficient manner. Specifically, the proposed sample retrieval rule aims to obtain more important samples when the noises are heteroscedastic. Due to the black-box nature of the pre-trained model, the local-smoothing step ensures the robustness of our method against adversarial pre-trained models. By selecting the optimal tuning parameters in Step 3, the proposed method automatically adapts to the best usage of the pre-trained model.

We will provide details of each step in the following sections.
In Sections~\ref{sec-tf} and~\ref{sec-adapt}, we present the bias correction and adaptation steps under a generic sampling scheme. Building on the resulting risk analysis, we then introduce the proposed sample retrieval rule in Section~\ref{sec-sample}, which approximates the optimal sampling strategy. The complete algorithm is summarized in Algorithm~\ref{alg2} at the end of this section.

\subsection{Smoothed bias correction}
\label{sec-tf}
Suppose that we have obtained $n$ labeled samples $\{(\bx_i^{\top},y_i)\}_{i=1}^n$, where the covariates $\bx_i$ are selected according to a pre-specified sampling scheme and the responses $y_i$ are generated from the target model \eqref{m1}. In this subsection, we introduce Step~2 of the proposed few-shot personalization framework.

As discussed earlier, the pre-trained model $f^{{(\ptr})}$ may be irregular or poorly aligned with the target regression function. To ensure stability of the subsequent bias correction, we first regularize $f^{{\ptr})}$ through a local smoothing operation.

For any $\bx^*,\bx'\in \calX$ and a given function $g(\cdot)$, define
\begin{align}
   & \omega_{\btheta,\bx^*}\circ g(\bx)=g(\bx^*)+ \min(|g(\bx)-g(\bx^*)|,\theta_1\|\bx-\bx^*\|_2^{\theta_2})\text{sgn}(g(\bx)-g(\bx^*)),\label{eq-tf}
\end{align}
where $\text{sgn}(\cdot)$ denotes the sign function.
We call this operation ``$\bm\theta$-local-smoothing''.
Loosely speaking, the transformation $\omega_{\btheta,\bx^*}(\cdot)$ enforces local H\"older regularity around $\bx^*$ by truncating excessive local variation in $g(\cdot)$. Importantly, the transformation is deterministic and does not depend on the observed data.

To formalize the resulting smoothness property, we introduce the following definition.
\begin{definition}[Local smoothness]
\label{def-ls}
For any $f(\cdot)$ defined on $\calX$ and any given $\bx\in\calX$, we say that $f\in L_{\bx}(\btheta)$ if
\[
  \sup_{\bx'\in \calX}\frac{|f(\bx')-f(\bx)|}{\|\bx'-\bx\|_2^{\theta_2}}\leq \theta_1
\]
for some finite $\theta_1>0$ and $\theta_2\geq 0$.
\end{definition}

The notion of local smoothness in Definition~\ref{def-ls} is weaker than global H\"older smoothness.  In particular, if $f\in H(\boldsymbol{\theta})$ with $0<\theta_2\le1$, then $f\in L_{\bx}(\boldsymbol{\theta})$ for all $\bx\in\mathcal{X}$. Therefore, if $g(\cdot)$ already satisfies a H\"older  condition with parameter $\btheta$, then $\omega_{\btheta,\bx^*}\circ g = g$. Conversely, local smoothness does not impose any global regularity away from the reference point $\bx^*$. 
By construction, for any given function $g(\cdot)$ defined on $\calX$, the transformed function $\omega_{\btheta,\bx}\circ g$ belongs to $L_{\bx}(\btheta)$ and satisfies $\omega_{\btheta,\bx^*}\circ g(\bx^*)=g(\bx^*)$ for any given $\bx^*$.

\begin{figure}[H]
\includegraphics[height=5.5cm,width=0.6\textwidth]{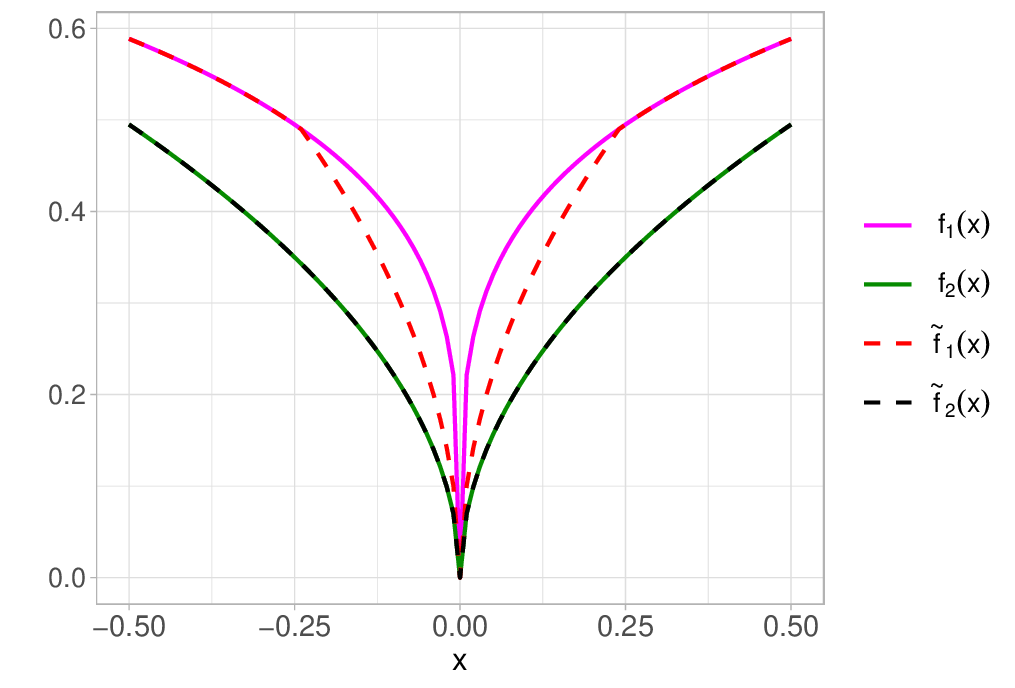}
\centering
\caption{Illustration of $\btheta$-local-smoothing. We set $f_1(x)=0.7|x|^{1/4}$ and $f_2(x)=0.7|x|^{1/2}$ for $x\in[-0.5,0.5]$.  Dashed lines correspond to $\tilde{f}_1(\bx)=\omega_{\btheta,0}\circ f_1(x)$ and $\tilde{f}_2(\bx)=\omega_{\btheta,0}\circ f_2(x)$ realized via (\ref{eq-tf}) with $\btheta=(1,0.5)^{\top}$.
}
\label{fig-tf}
\end{figure}

An illustration is given in Figure \ref{fig-tf} with $\btheta=(1,0.5)^{\top}$.  We see that function $f_1(x)=0.7|x|^{1/4}\not\in L_0(\btheta)$ for $x\in[-0.5,0.5]$ and it is transformed to $\tilde{f}_1(\bx)=\omega_{\btheta,0}\circ f_1(\bx)$ which is flatter around zero. Function $f_2(x)=0.7|x|^{1/2}\in H(\btheta)$ and it is unchanged by (\ref{eq-tf}).  We will apply this transformation to $f^{(\ptr)}(\cdot)$ and show that it can lead to the desired theoretical performance for our final estimator.

Next, we introduce the bias correction step. Given $n$ retrieved samples, we use a large proportion to estimate the model and use a small fraction as validation samples for selecting the tuning parameters. The sample splitting step is described in Section \ref{sec-sample}. Let $\mathcal{N}^{(tr)}\subseteq[n]$ denote the index set of training samples and assume it is given for now. 
For any given $\bx\in\calX$, let
   \begin{align}
\hat{\delta}_{\btheta}(\bx)&=\frac{\sum_{i\in\mathcal{N}^{(tr)}}(y_i-\omega_{\btheta,\bx}\circ f^{(\ptr)}(\bx_i))\mathbbm{1}(\|\bx_i-\bx\|_{\infty}\leq h)}{1\vee \sum_{i\in\mathcal{N}^{(tr)}}\mathbbm{1}(\|\bx_i-\bx\|_{\infty}\leq h)}, \label{eq-hdelta}
\end{align}
The estimator $\hat{\delta}_{\btheta}(\bx)$ corresponds to a kernel estimator of the bias function $\delta_{\btheta,\bx}(\bx)=f^*(\bx)-\omega_{\btheta,\bx}\circ f^{(\ptr)}(\bx)$, exploiting the local smoothness of $\delta_{\btheta,\bx}(\cdot)$ around $\bx$. 
As  equation (\ref{eq-tf}) implies that $\omega_{\theta,\bx}\circ f^{(\ptr)}(\bx)=f^{(\ptr)}(\bx)$, the estimator $\hat{\delta}_{\btheta}(\bx)$ is also a proper estimate of $f^*(\bx)-f^{(\ptr)}(\bx)$. 
Hence, we define the calibrated estimator as
\begin{align}
   \hat{f}^{(\fsp)}_{\btheta}(\bx)&= f^{(\ptr)}(\bx)+ \hat{\delta}_{\btheta}(\bx),~\bx\in\calX.\label{eq-fsp}
   \end{align}
In view of (\ref{eq-hdelta}) and (\ref{eq-fsp}), the computation of $\hat{f}^{(\fsp)}_{\btheta}(\bx)$, for any given $\btheta$ and $\bx\in\calX$, only need to query $f^{(\ptr)}$ at the retrieved $\bx_i,i\in\mathcal{N}^{(tr)}$ and each test point, which enjoys computational efficiency.
   
   Note that when we apply $\bm\theta$-local-smoothing with  $\theta_1=0$, $\omega_{\btheta,\bx}\circ f^{(\ptr)}(\bx')=f^{(\ptr)}(\bx)$ for any $\bx'\in\calX$ and hence $\hat{f}^{(\fsp)}_{\btheta}(\cdot)$ reduces to the single-task kernel estimate only based on the target samples. In this case ($\theta_1=0$), the pre-trained model is not leveraged in the personalized estimate.
 For $\theta_1>0$, information from the pre-trained model is incorporated in a controlled manner through local smoothing. Consequently, the tuning parameter ${\btheta}$ governs the extent to which the pre-trained model influences the final estimator, ensuring robustness even when $f^{({\ptr})}$ is poorly aligned with $f^*$. 
   
Next, we will leverage validation samples to choose a proper $\btheta$ as detailed in the next subsection. 

\subsection{Adaptation}
\label{sec-adapt}

We leverage the cross-validation technique to select tuning parameters $\btheta$.
Let 
$\Theta=\{0,c_1/\log n,2c_1/\log n,\dots,c_1\}\times \{0,1/\log n,2/\log n,\dots,1\}$ be a grid for search optimal $\btheta$, where $c_1$ is a pre-determined constant. We first fit $\{\hat{f}^{(\fsp)}_{\btheta}\}_{\btheta\in\Theta}$ based on the training samples $\mathcal{N}^{(tr)}$. Let $\mathcal{N}^{(va)}\subseteq[n]\setminus\mathcal{N}^{({tr)}}$ denote set the validation samples.
Let us assume $\mathcal{N}^{(tr)}$ and $\mathcal{N}^{(va)}$ are given for now and we will discuss how to split $n$ retrieved samples into training and validation samples in the next subsection.
Specifically, let
\begin{align}
\label{eq-cv}
\hat{\btheta}&=\argmin_{\btheta\in\Theta} \sum_{i\in\mathcal{N}^{(va)}} (y_i-\hat{f}^{(\fsp)}_{\btheta}(\bx_i))^2.
\end{align}
That is, we choose $\hat{\btheta}$ to be the best tuning parameter that minimizes the mean prediction error in the validation samples.
Including candidate values with $\theta_1=0$ in the grid $\Theta$ ensures adaptivity to the informativeness of the pre-trained model. In particular, when the pre-trained model provides little or no useful signal for the target task, the procedure can select a tuning parameter that effectively ignores $f^{(\ptr)}$, thereby reducing to a target-only estimator.


\subsection{Sample retrieval scheme}
\label{sec-sample}
In this section, we introduce the proposed retrieval scheme and give details about the sample splitting. To motivate its design, we first analyze the theoretical properties of the estimator $\hat{f}^{(\fsp)}_{\btheta}(\bx)$ for a fixed tuning parameter $\btheta$ under a generic pre-determined sampling scheme $p_X(\bx)$ on $\mathcal{X}$. The analysis highlights how the sampling distribution affects the mean integrated squared error (MISE) and guides the construction of an efficient retrieval strategy.

We begin by stating standard regularity conditions for nonparametric regression.
\begin{condition}[Global smoothness]
\label{cond1}
Assume that $f^*(\bx)\in H(\btheta^*)$ on $\mathcal{X}$ for some finite $\theta^*_1\geq 0$ and $0<\theta^*_2\leq 1$. 
\end{condition}

\begin{condition}[Sub-Gaussian noise]
\label{cond-noise}
Assume that the $\eps_i$ are independent sub-Gaussian with mean zero and variance $\sig^2(\bx_i)$ for $i=1,\dots,n$. There exists some positive constant $\sig_{\max}$ such that $\sup_{\bx\in\mathcal{X}}\sig^2(\bx)\leq \sig^2_{\max}$.  The function $\sig(\bx)\in H(\btheta^{(\sig)})$ on $\mathcal{X}$ for some finite $\theta_1^{(\sig)}\geq 0$ and $\theta^{(\sig)}_2>0$.
\end{condition}

Condition \ref{cond1} specifies the H\"older smoothness of $f^*(\cdot)$. As discussed before, this is a standard assumption in transfer learning and in classical nonparametric literature. Condition~\ref{cond-noise} allows for heteroskedastic sub-Gaussian noises and imposes mild smoothness on the variance function, which facilitates variance estimation.

In the next lemma, we first demonstrate the effect of the sampling distribution $p_X(\bx)$ on the MISE as a motivation for the proposed retrieval scheme. Since the pre-trained model $f^{(\ptr)}(\cdot)$ is learned from an external data source independent of the retrieved samples, we treat it as deterministic in the analysis without loss of generality.

\begin{lemma}[MISE under generic retrieval scheme]
\label{tlem1}
Assume that Conditions \ref{cond1} and \ref{cond-noise} hold true. Suppose that $\bx_i$ are generated according to some pre-deterimined $p_X(\bx)$ and the corresponding $y_i$ are generated according to (\ref{m1}). Then for any finite $\theta_1\geq 0$ and any $0<\theta_2\leq 1$, we have
\begin{align*}
\textup{MISE}(\hat{f}^{(\fsp)}_{\btheta})&\lesssim \gamma^2_1(\btheta)(\sqrt{d}h)^{2\gamma_2(\btheta)}+\int_{\calX}\E[\frac{\sig^2(\bx)+\sig_{\max}\theta_1^{(\sig)}h^{\theta_2^{(\sig)}}}{n_h(\bx)\vee 1}]d\bx,
\end{align*}
where $n_h(\bx)=\sum_{i=1}^n\mathbbm{1}(\|\bx_i-\bx\|_{\infty}\leq h)$  and $\bm\gamma(\btheta)$ are the smoothness parameters such that $\delta_{\btheta,\bx}\in L_{\bx}(\bm\gamma(\btheta))$ for all $\bx\in\calX$.
\end{lemma}
The first term in the upper bound of Lemma \ref{tlem1} corresponds to the squared bias of the estimator and depends on the local smoothness of the bias function $\delta_{\btheta}(\bx)$ after local smoothing. We will discuss the magnitude of $\gamma(\btheta)$ in the next section and focus on the variance reduction effects by sample retrieval in the rest of this section. The second term in the upper bound captures the variance contribution, which is inversely proportional to the number of local samples $n_h(\bx)$. Under mild conditions on the bandwidth and retrieval distribution, the term $\sig_{\max}\theta_1^{(\sig)}h^{\theta_2^{(\sig)}}$ is negligible and $n_h(\bx)\propto nh^dp_X(\bx)$. Consequently, the ideal choice of $p_X(\bx)$ should minimize $\int_{\bx\in \calX}\sig^2(\bx)/p_X(\bx)d\bx$, suggesting that the optimal sampling density should allocate more samples to regions with higher noise levels.  This leads to the following optimal sampling distribution:
\begin{align}
\label{eq-nstar}
p^*_X(\bx)=\frac{\sig(\bx)}{\int_{\calX}\sig(\bx)d\bx},~\bx\in\calX,
\end{align}
This form aligns with the intuition that regions with larger noise variance require denser sampling for accurate estimation. 
We next derive the corresponding MISE bound under the optimal retrieval scheme. Let \[
\bar{\sig}=\frac{\int_{\calX}\sig(\bx)d\bx}{|\calX|}
\]
 denote the average noise level. For a given bandwidth $h$, define $r_n=\max\{\sqrt{\sig_{\max}\theta_1^{(\sig)}h^{\theta_2^{(\sig)}}},\frac{\sig_{\max}\log n}{nh^d}\}$ and $\calX_0=\left\{\bx\in\mathcal{X}:\sig(\bx)\leq c_rr_n\right\}$  for a large enough constant $c_r$.
\begin{corollary}[MISE under optimal retrieval scheme]
\label{cor1}
Assume that Conditions \ref{cond1} and \ref{cond-noise} hold. Suppose that $\bx_i$, $i=1,\dots,n$, are generated according to  $p^*_X(\bx)$ and the corresponding $y_i$ according to (\ref{m1}).
If $h=n^{-c_0}$ for some constant $c_0\geq 0$ and $|\calX_0|\leq c_1\min\{\bar{\sig}^2/(r_n^2nh^d),1\}$ for some constant $c_1<1$, then for any finite $\theta_1\geq 0$ and $\theta_2\geq 0$, we have
\begin{align}
\label{re1-thm}
\textup{MISE}(\hat{f}^{(\fsp)}_{\btheta})&\lesssim \gamma^2_1(\btheta)(\sqrt{d}h)^{2\gamma_2(\btheta)}+\frac{\bar{\sig}^2}{nh^d},
\end{align}
where $\bm\gamma(\btheta)$ are the smoothness parameters such that $\delta_{\btheta,\bx}(\cdot)\in L_{\bx}(\bm\gamma(\btheta))$ for all $\bx\in\mathcal{X}$.
\end{corollary}



%

Corollary \ref{cor1} establishes the convergence rate of the personalized estimate under the optimal retrieval scheme $p^*_X(\cdot)$. The first term on the right hand side of (\ref{re1-thm}) corresponds to the squared bias, while the second term represents the variance contribution.  The condition in Corollary~\ref{cor1} requiring $|\mathcal{X}_0|$ to be sufficiently small ensures that each local kernel neighborhood contains enough samples. Under the optimal sampling scheme, the effective noise level of the estimator is characterized by the average variance $\bar{\sig}^2$. The resulting variance reduction effect is further illustrated in Remark~\ref{re2}.

\begin{remark}[Variance comparison]
\label{re2}
Consider the uniformly sampling scheme where we randomly sample $n$ points from $\calX=[0,1]$. Suppose that $\sig(\bx)=b_n-x$ for $x\in[0,b_n-a_n]$ and $\sig(\bx)=a_n$ for $\bx\in [b_n-a_n,1]$ for $b_n>a_n$.  The variance term under this uniformly random sampling scheme is
\begin{align*}
&\frac{\int_{\calX}\sig^2(\bx)d\bx}{nh^d}\geq \frac{\bar{\sig}^2}{nh^d}+\frac{(b_n-a_n)^3}{12}.
\end{align*}
The last term shows the efficiency loss of randomly sampling over the optimal sampling scheme when $a_n\neq b_n$.

If $b_n-a_n=o(1)$ and $a_n=o(b_n-a_n)$, then
\begin{align}
\label{eq1-re2}
\frac{\int_{\bx\in \calX}\sig^2(\bx)d\bx}{nh^d}\gg \frac{\bar{\sig}^2}{nh^d},
\end{align}
implying that the optimal sampling scheme yields a strictly smaller variance order than uniform sampling.
\end{remark}

The proof of Remark \ref{re2} is defered to the supplement.
Remark \ref{re2} demonstrates that when the noise is heteroskedastic, the optimal sampling scheme can substantially reduce the variance of the estimator relative to uniform sampling. In particular, Equation \eqref{eq1-re2} shows that the variance order itself can be improved when the noise variance is small over a sufficiently large region.
A canonical example where the noise is heteroskedastic is in classification tasks. 
Concretely, if $\P(y_i=1|\bx_i)=f^*(\bx_i)$, then $\sig^2(\bx)=f^*(\bx)(1-f^*(\bx))$. Consequently, the noise variance is small in regions where $f^*(\bx)$ is close to $0$ or $1$, and largest when $f^*(\bx)$ is near $1/2$, making adaptive sampling particularly advantageous.

While the optimal density $p_X^*(\bx)$ depends on the unknown variance function $\sigma^2(\bx)$, it can be approximated from data. To this end, we allocate a small fraction of the sampling budget to estimate $\sigma^2(\bx)$ and use the resulting estimate to guide sample retrieval. The procedure is summarized in Algorithm~\ref{alg-sample1}.

\begin{algorithm}[H]
\begin{flushleft}
\textbf{Input}: Sampling budget $n$ and target region $\calX$.

\textbf{Output}:  Retrieved samples $(\bx_i^{\top},y_i)$, $i=1,\dots,n$, $\mathcal{N}^{(tr)}$, and $\mathcal{N}^{(va)}$.

\textbf{Step 1.} For some small constant $c$, randomly retrieve $n_0=cn$ labeled samples $\{\bx_i,y_i\}_{i=1}^{n_0}$ from $\calX$. For $h_{\sig}=n^{-1/(d+2)}$ and $\bx\in\mathcal{X}$, compute
\begin{align}
\hat{\sigma}^2(\bx)=\frac{\sum_{i=1}^{n_0/2} y_i^2K_{h_{\sig}}(\bx_i,\bx)}{1\vee\sum_{i=1}^{n_0/2}K_{h_{\sig}}(\bx_i,\bx)}-\frac{\{\sum_{i=1}^{n_0/2} y_iK_{h_{\sig}}(\bx_i,\bx)\}^2}{1\vee \{\sum_{i=1}^{n_0/2}K_{h_{\sig}}(\bx_i,\bx)\}^2},\label{eq-hsig}
\end{align}
where $K_h(\bx,\bx')=\max\{0,h-\|\bx-\bx'\|_{\infty}\}$.

 \textbf{Step 2}. 
Sample $\bx_i$, $i=n_0+1,\dots, n$, from 
\[\hat{p}_X(\bx)=\frac{\hat{\sig}(\bx)}{\int_{\calX} \hat{\sig}(\bx)d\bx}
\]

Set $\mathcal{N}^{(tr)}=\{n_0+1,\dots,n\}$ and $\mathcal{N}^{(va)}=\{n_0/2+1,\dots,n_0\}$. Output  $\bx_i$ and their corresponding responses $y_i$, $i=1,\dots, n$.
\end{flushleft}
\caption{Sample retrieval scheme.}
\label{alg-sample1}
\end{algorithm}

We now provide more illustrations of Algorithm \ref{alg-sample1}. In Step 1, we first retrieve $n_0$ samples uniformly random from $\calX$. Half of the samples are used to compute $\hat{\sig}^2(\bx)$ and the other half will be used for validation in the adaptation step. Once the variance function is estimated, the proposed retrieved samples $\hat{p}_X(\cdot)$ can be estimated. We estimate $\sig^2(\bx)$ based on a kernel estimate. The kernel estimator \eqref{eq-hsig} is computationally simple, does not require prior knowledge of smoothness parameters and is consistent under mild conditions. In Step 2, sampling from $\hat{p}_X(\bx)$ can be implemented using standard rejection sampling methods \cite{casella2004generalized,gilks1992adaptive}. For completeness, the detailed algorithm for rejection sampling is given by Algorithm A.1 in the supplement.

 Note that our choice for the bandwidth $h_{\sig}$ is for simplicity. Indeed, as long as the width of $h_{\sig}$ is $o(1)$ and the number of samples for each local estimate grows to infinity, the estimate $\hat{\sig}^2(\bx)$ is consistent and the theoretical guarantees in next subsection still hold.

In the next lemma, we justify the performance of the proposed retrieval scheme. Let $\calX_1=\{x\in\calX:\sig(\bx)\leq c_r\tilde{r}_n\}$ for $\tilde{r}_n=r_n+\log n\cdot n^{-\frac{\min(2\theta^*_2,\theta^{(\sig)}_2,1/2)}{4+2d}}$ for a sufficiently large constant $c_r$.   
\begin{lemma}[Rate optimality of the proposed retrieval scheme]
\label{lem-retrieval}
Assume Conditions \ref{cond1}, \ref{cond-noise}. 
Suppose that $h=n^{-c_0}$ for some constant $c_0\geq 0$ and $|\calX_1|\leq c_1\min\{\bar{\sig}^2/(\tilde{r}^2_nnh^d),1\}$ for some positive constant $c_1<1$. Then there exists some positive constant $C$ such that
\[
\int_{ \calX}\E[\frac{\sig^2(\bx)+\sig_{\max}\theta_1^{(\sig)}h^{\theta_2^{(\sig)}}}{\max\{1,n_h(\bx)\}}]d\bx\leq C\frac{\bar{\sig}^2}{nh^d}.
\]
\end{lemma}

Lemma \ref{lem-retrieval} shows that the variance term under the proposed retrieval scheme $\hat{p}_X(\cdot)$ achieves the same rate as the optimal retrieval scheme $p^*_X(\cdot)$. This result demonstrates the effectiveness of the proposed retrieval scheme. The additional restriction on $|\mathcal{X}_1|$ is used to control the error incurred when estimating $\sigma(\bx)$. 
Finally, we summarize the full personalization algorithm combining the sampling phase and the learning phase in Algorithm \ref{alg2}.
\begin{algorithm}[H]
\begin{flushleft}
\textbf{Input}:  Pre-trained model $f^{(\ptr)}$, sampling budget $n$, and target region $\calX$.

\textbf{Output}: $\hat{f}^{(\fsp)}_{\hat{\btheta}}(\bx)$, $\bx\in\mathcal{X}$.
 
\textbf{Step 1}. Obtain $\hat{p}_X(\bx)$ and generate samples $(\bx_i^{\top},y_i)$, $i=1,\dots,n$ by Algorithm \ref{alg-sample1}. 

\textbf{Step 2}. For each $\btheta\in\Theta$, compute $\hat{f}^{(\fsp)}_{\btheta}(\bx)$  via (\ref{eq-fsp})  with $\mathcal{N}^{(tr)}=\{n_0+1,\dots,n\}$.  

\textbf{Step 3}. Estimate model weights $\hat{\btheta}$ via (\ref{eq-cv}) with $\mathcal{N}^{(va)}=\{n_0/2+1,\dots,n_0\}$, and output $\hat{f}_{\hat{\btheta}}^{(\fsp)}(\bx)$.
\end{flushleft}
\caption{Personalized nonparametric estimate of $f^*(\bx)$, $\bx\in\mathcal{X}$.}
\label{alg2}
\end{algorithm}

The bandwidth $h$ can also be selected via cross-validation. For instance, one may search jointly over $(\boldsymbol{\theta},h)\in\Theta\times\mathcal{H}$ with $\mathcal{H}=\{1,1/2,\ldots,1/n\}$. The corresponding theoretical analysis follows similar arguments to \eqref{eq-cv} and is omitted for brevity.


\section{Theoretical properties of Algorithm \ref{alg2}}
\label{sec-theory}

In this section, we establish theoretical guarantees for the proposed few-shot personalization procedure in Algorithm \ref{alg2}.

We first upper bound the risk of the personalized estimator under the proposed retrieval scheme.

\begin{theorem}[MISE under proposed retrieval scheme]
\label{thm-final}
Assume Conditions \ref{cond1} and \ref{cond-noise} hold. 
 Suppose that $h=n^{-c_0}$ for some constant $c_0\geq 0$ and $|\calX_1|\leq c_1\min\{\bar{\sig}^2/(\tilde{r}^2_nnh^d),1\}$ for some positive constant $c_1<1$. Then
\begin{align*}
\textup{MISE}(\hat{f}_{\btheta}^{(\fsp)})\lesssim\gamma_1^2(\btheta)h^{2\gamma_2(\btheta)}+\frac{\bar{\sig}^2}{nh^d},
\end{align*}
where $\bm\gamma(\btheta)$ are the smoothness parameters such that $\delta_{\btheta,\bx}(\cdot)\in L_{\bx}(\bm\gamma(\btheta))$ for all $\bx\in\mathcal{X}$.

If we take $h\asymp \min\{(\bar{\sig}/\gamma_1(\btheta))^{\frac{2}{2\gamma_2(\btheta)+d}} n^{-\frac{1}{2\gamma_2(\btheta)+d}},1\}$, then
\begin{align}
\textup{MISE}(\hat{f}_{\btheta}^{(\fsp)})&\lesssim  (\gamma_1(\btheta))^{\frac{2d}{2\gamma_2(\btheta)+d}}\bar{\sig}^{\frac{4\gamma_2(\btheta)}{2\gamma_2(\btheta)+d}} n^{-\frac{2\gamma_2(\btheta)}{2\gamma_2(\btheta)+d}}+\frac{\bar{\sig}^2}{n}.\label{re-thm-final}
\end{align}
\end{theorem}
Theorem~\ref{thm-final} provides an MISE upper bound for $\hat{f}_{\btheta}^{(\fsp)}(\bx)$ under the proposed data-driven retrieval scheme for fixed $\btheta$. 
The bound is comparable to Corollary~\ref{cor1}, showing that the retrieval procedure achieves the same variance order as the oracle sampling density. By taking the bandwith $h$ to balance the bias and variance term, we obtain the upper bound (\ref{re-thm-final}), which will be shown to be optimal  in Theorem \ref{thm-mini1}. 

The bound \eqref{re-thm-final} contains a nonparametric term and a parametric term. The nonparametric term arises from balancing the bias and variance in estimating ${\delta}_{\btheta}(\bx)$ via \eqref{eq-hdelta}. quantity related to $n$, $n^{-\frac{2\gamma_2(\btheta)}{2\gamma_2(\btheta)+d}}$, is the usual minimax optimal rate for estimating a nonparametric function with H\"older smoothness $\gamma_2(\btheta)$. In contrast to classical analyses that treat the H\"older constant as fixed, our bound tracks the dependence on $\gamma_1(\btheta)$, which in the personalization setting may be small.
The second term of (\ref{re-thm-final}) is a parametric rate. It becomes dominant when $\gamma_1(\btheta)$ is close to zero, in which case the $\delta_{\btheta}(\bx)$ is close to a constant. 
Correspondingly, the optimal rate for estimating an unknown constant based on $n$ samples with noise level $\sig(\bx_i)$ is $\bar{\sig}^2/n$, which corresponds to the second term. As $\gamma_1(\btheta)$ can be very small and the width of $\calX$ is constant, we take $h$ to be no larger than a constant, which leads to the parametric term in the upper bound.

We next discuss how the local smoothing tuning parameter $\btheta$  in step (\ref{eq-tf}) affects the local smoothness parameters $\bm\gamma(\btheta)$. 
Since $f^*(\cdot)\in H(\btheta)$, it follows that
$\gamma_2(\btheta)\geq \min\{\theta^*_2,\theta_2\}$ and $\gamma_1(\btheta)\leq \theta_1+\theta^*_1$. In particular, choosing $\theta_2=\theta^*_2$ ensures $\gamma_2(\btheta)\geq \theta_2^*$. In many settings, $\gamma_1(\btheta)$ can be substantially smaller than $\theta^*_1$ and $\gamma_2(\btheta)$ can be larger than $\theta_2$, reflecting reduced local H\"older complexity  after incorporating the pre-trained model. The following examples illustrate these effects.

\begin{example}
\label{ex1}
Consider $f^*(\bx)=f_1^*(\bx)+f_2^*(\bx)$, where $f_1^*\in H(\btheta)$  and $f_2^*\in H(\btheta')$ for $\theta_2<\theta'_2$. Suppose that $f^{(ptr)}=f_1^*$.
Then $f^*(\bx)\in H(\btheta^*)$ for $\btheta^*=(\theta_1+\theta_1',\theta_2)$. Moreover, by setting $\btheta=\btheta^*$, $\omega_{\btheta,\bx}\circ f^{(\ptr)}(\cdot)=f^{(\ptr)}(\cdot)$ for all $\bx$ and $\delta_{\btheta^*,\bx}(\cdot)=f_2^*(\cdot)\in H(\btheta')$. As $\theta_2<\theta'_2$, it shows that $\delta_{\btheta^*,\bx}(\cdot)$ is smoother than $f^*(\cdot)$ in the sense of local smoothness characterization.
\end{example}

\begin{example}
\label{ex2}
Consider  $f^*\in H(\btheta^*)$ and $f^{(ptr)}(\bx)=\rho f^*(\bx)$ for some constant $0<\rho<1$.
Then for $\btheta=(\rho\theta^*_1,\theta^*_2)$, we have  $\omega_{\btheta,\bx}\circ f^{(\ptr)}(\cdot)=f^{(\ptr)}(\cdot)$ for all $\bx$ and $\delta_{\btheta^*,\bx}(\cdot)=(1-\rho)f^*(\cdot)\in H(\btheta')$
 for $\btheta'=((1-\rho)\theta^*_1,\theta^*_2)$. This is an example of $\gamma_1(\btheta)<\theta^*_1$.
\end{example}

In view of above two examples, appropriate choices of $\btheta$ in (\ref{eq-tf}) can lead to higher local smoothness of $\delta_{\btheta,\bx}(\cdot)$, which can improve the rate in \eqref{re-thm-final}. 
We now compare the bound in (\ref{re-thm-final}) to the MISE of conventional single-task nonparametric estimates. If $\gamma_1(\btheta)$ and $\bar{\sigma}$ are both constants bounded away from 0, then MISE$(\hat{f}_{\btheta}^{(\fsp)})$ is of order $n^{-\frac{2\gamma_2(\btheta)}{2\gamma_2(\btheta)+d}}$. In comparison, if we estimate $f^*(\cdot)$ only based on $n$ target samples, then the MISE is of order $n^{-\frac{2\theta^*_2}{2\theta^*_2+d}}$. To guarantee that $\hat{f}_{\btheta}^{(\fsp)}$ is no worse in comparison to the conventional single-task nonparametric estimate, the tuning parameter $\btheta$ in (\ref{eq-tf}) need to be chosen properly. Specifically, setting $\theta_2=\theta^*_2$ guarantees $\gamma_2(\btheta)\geq \theta_2$. If $\gamma_2(\btheta)$ is strictly larger than $\theta_2$ as in Example \ref{ex1}, then $\hat{f}_{\btheta}^{(\fsp)}$ has a faster convergence rate than the single-task estimate, which demonstrates the benefits of leveraging the pre-trained model.
Another potential gain of $f^{(\fsp)}_{\btheta}$ is that $\gamma_1(\btheta)$ can be much smaller than $\theta_1$ as illustrated in Example \ref{ex2}. In this case, the MISE of $\hat{f}^{(\fsp)}_{\btheta}$ can also have a smaller order than the conventional single-task nonparametric estimator. To summarize, by choosing $\btheta$ property,  our proposal can have a faster convergence rate by improving the H\"older smoothness.

We now justify the effectiveness of the adaptation step.

\begin{theorem}[MISE after adaptation]
\label{thm-adapt}
Assume the conditions of Theorem \ref{thm-final}. For $\hat{\btheta}$ obtained from (\ref{eq-cv}), it holds that
\begin{align*}
\textup{MISE}(\hat{f}^{(\fsp)}_{\hat{\btheta}})\leq C_1\min_{\btheta\in\Theta}\textup{MISE}(\hat{f}_{\btheta}^{(\fsp)})+\frac{C_2\sig^2_{\max}(\log n)^2}{n}
\end{align*}
for some positive constants $C_1$ and $C_2$.
\end{theorem}
Theorem~\ref{thm-adapt} shows that validation step selects a tuning parameter achieving the best risk over the candidate set $\Theta$ up to a small remainder term. The remainder term, reflecting the cost of adaptation, is of order $(\log n)^2/n$, where $(\log n)^2$ is indeed the cardinality of the search space $\Theta$, $|\Theta|$. This term is negligible relative to the nonparametric term in \eqref{re-thm-final} in the regimes of primary interest. The proposed tuning parameter selection step is simple because $|\Theta|$ is small in the current setting.
If $|\Theta|$ is large, more advanced model selection methods can be leveraged \cite{dai2012deviation,lecue2014optimal}.

Next, we establish the minimax optimality of the proposed method.
We define the following functional space
\begin{align}
\mathcal{F}_{\btheta}(\btheta^*,\bm\gamma,\sigma_0)&=\left\{P_{y|\bx}:f^*(\bx)\in H(\btheta^*)~\forall~\bx\in\mathcal{X},~\left(\int_\mathcal{X}\sig(\bx)d\bx\right)/|\calX|\leq \sig_0,\right.\nonumber\\
&\quad\quad\left. f^*(\cdot)-\omega_{\btheta,\bx}\circ f^{(\ptr)}(\cdot)\in L_{\bx}(\bm\gamma)~\forall \bx\in\calX\right\},\label{space1}
\end{align}
where $\btheta$ denotes a pre-determined tuning parameter in the local smoothing step and $P_{y|\bx}$ denotes the conditional distribution of $y$ given $\bx$.
This functional space considers H\"older smooth target function $f^*(\cdot)$ with smoothness parameter $\btheta^*$. Moreover, the parameter $\bm\gamma$ denotes the local smoothness parameter of $\delta_{\btheta,\bx}(\cdot)$, the bias function after $\btheta$-local-smoothing. In the next theorem, we establish the minimax rate for MISE in the functional space (\ref{space1}). 
\begin{theorem}[Minimax lower bound]
\label{thm-mini1}
Let $p_X(\bx)$ denote the sampling distribution for generating retrieved covariates $\bx_i$, $i=1,\dots,n$ and $\calD_n=\{\bx_i^{\top},y_i\}_{i=1}^n$ denote the retrieved samples. For $0\leq \theta_1\leq C_1\theta^*_1$ and $\theta_2\geq \theta^*_2$, there exists some positive constant $C_2$ such that
\begin{align}
\label{eq-lb1}
\inf_{ p_X,\hat{f}(\calD_n,f^{(\textup{ptr})})}\sup_{f^*\in\mathcal{F}_{\btheta}(\btheta^*,\bm\gamma,\sig_0)}\textup{MISE}(\hat{f})\geq C_2\gamma_1^{\frac{2d}{2\gamma_2+d}}\sig_0^{\frac{4\gamma_2}{2\gamma_2+d}} n^{-\frac{2\gamma_2}{2\gamma_2+d}}+C_2\frac{\sig_0^2}{n},
\end{align}
where $\gamma_1\leq C_3\theta^*_1$ and $\gamma_2\geq \theta^*_2$ for some positive constant $C_3$.
\end{theorem}
Theorem~\ref{thm-mini1} provides a minimax lower bound over $\mathcal{F}_{\btheta}(\btheta^*,\bm\gamma,\sigma_0)$ and takes the infimum over both the retrieval distribution $p_X$ and all estimators based on $(\mathcal{D}_n,f^{({\ptr})})$. This formulation captures the effect of user-specified retrieval schemes on the minimax risk. Under the stated conditions on ${\btheta}$, the local regularity parameters satisfy $\gamma_1=O(\theta_1^*)$ and $\gamma_2\ge \theta_2^*$, implying that the minimax personalization rate is no worse than the classical single-task minimax rate.


Theorems~\ref{thm-final} and \ref{thm-mini1} together show that, for a fixed tuning parameter $\btheta$, the estimator $\hat{f}^{(\fsp)}_{\btheta}(\bx)$ attains the minimax optimal rate. Moreover, by Theorem~\ref{thm-adapt}, the estimator $\hat{f}^{(\fsp)}_{\hat{\btheta}}(\bx)$ achieves the oracle risk over $\Theta$ up to logarithm factors. As commented above, the logarithm inflation is negligible in common scenarios because the first term on the right-hand side of (\ref{eq-lb1}) is nonparametric rate. 


\section{Extensions}
\label{sec-xstar}
In this section, we consider extensions of the proposed algorithm to two additional scenarios.
\subsection{Extension to infinitesmall $\calX$}
\label{sec1-xstar}
In some applications, the target covariate region $\mathcal{X}$ may be small, rather than having constant Lebesgue measure. For example, one may wish to personalize a generic prediction model to a small subpopulation, in which case the target covariate support may satisfy $|\calX|=o(1)$. We study this regime by considering $\calX=[0,\nu_n]^d$ for $\nu_n\leq 1$ and $\nu_n$ is allowed to go to zero as $n$ goes to infinity.

When $\nu_n$ is small, the variance-optimizing weighted sampling scheme becomes less critical,  since the heterogeneity of $\sigma(\bx)$ over $\calX$ is limited by the diameter of the region. We therefore consider uniform sampling over $\calX$ and adapt Algorithm~\ref{alg2} accordingly.

\begin{algorithm}[H]
\begin{flushleft}
Input:  Pre-trained model $f^{(\ptr)}$, sampling budget $n$, and target region $\calX=[0,\nu_n]^d$.

 Output: $\hat{f}_{\hat{\btheta}}(\bx)$, $\bx\in\mathcal{X}$.
 
\textbf{Step 1}. Uniformly sample $\bx_i$, $i=1,\dots,n$ from $\calX$ and  their corresponding responses $y_i$ based on (\ref{m1}).

\textbf{Step 2}. For each $\btheta\in\Theta$, compute $\hat{f}^{(\fsp)}_{\btheta}(\bx)$  via (\ref{eq-fsp})  with $\mathcal{N}^{(tr)}=\{n_0+1,\dots,n\}$ and bandwith $h\leq \nu_n$.  

\textbf{Step 3}. Estimate model weights $\hat{\bm{\alpha}}$ via (\ref{eq-cv}) with $\mathcal{N}^{(va)}=\{1,\dots,n_0\}$, and output $\hat{f}_{\hat{\btheta}}^{(\fsp)}(\bx)$.
\end{flushleft}
\caption{Personalized nonparametric estimate of $f^*(\bx)$, $\bx\in\mathcal{X}$ for infinitesmall $\calX$.}
\label{alg-xstar}
\end{algorithm}

Algorithm~\ref{alg-xstar} differs from Algorithm~\ref{alg2} primarily in the retrieval step (uniform sampling on $\mathcal{X}$) and in the additional constraint $h\le \nu_n$, which ensures that the local neighborhoods used by the kernel estimator remain contained in the target region. The following corollary provides the corresponding risk bound.
\begin{corollary}
\label{cor-xstar}
Assume Conditions \ref{cond1}, \ref{cond-noise}. 
Suppose that  $\nu_n=n^{-c_0}$ for some constant $c_0\geq 0$ and $\bar{\sig}^2\geq \nu_n^{\theta_2^{(\sig)}}$. 
If we take $h\asymp \min\{\nu_n^{\frac{d}{2\gamma_2(\btheta)+d}}(\bar{\sig}/\gamma_1(\btheta))^{\frac{2}{2\gamma_2(\btheta)+d}} n^{-\frac{1}{2\gamma_2(\btheta)+d}},\nu_n\}$, then for any finite $\theta_1\geq 0$ and $\theta_2\geq 0$, we have
\begin{align}
\textup{MISE}(\hat{f}_{\btheta}^{(\fsp)})/\nu_n^d&\lesssim (\gamma_1(\btheta))^{\frac{2d}{2\gamma_2(\btheta)+d}}\nu_n^{\frac{2d\gamma_2(\btheta)}{2\gamma_2(\btheta)+d}}\bar{\sig}^{\frac{4\gamma_2(\btheta)}{2\gamma_2(\btheta)+d}} n^{-\frac{2\gamma_2(\btheta)}{2\gamma_2(\btheta)+d}}+\frac{\bar{\sig}^2}{n}.\label{re-cor2}
\end{align}
\end{corollary}

In (\ref{re-cor2}), we establish the upper bound for the MISE of the personalized estimator $\hat{f}_{\btheta}^{(\fsp)}(\cdot)$ on a shrinking domain for any given $\btheta$. Since the measure of $\calX$ is $\nu_n^d$, the quantity $\textup{MISE}(\hat{f}_{\btheta}^{(\fsp)})/\nu_n^d$ corresponds to average integrated mean squared error over $\mathcal{X}$. The bound shows that, holding other parameters fixed, the average error decreases as $\nu_n$ shrinks. Intuitively, a smaller domain permits a smaller effective bandwidth and reduces the difficulty of the nonparametric estimation problem. In this regime, uniform sampling achieves the same order as variance-weighted sampling, so explicit reweighting provides limited additional benefit.

 

\subsection{Existence of unlabelled data from $\calX$}

In some practical scenarios, it may not be possible to freely generate labeled samples from the target distribution. A common alternative setting for few-shot personalization is one in which a large collection of unlabeled covariates from the target region is available, but only a limited number of labels can be acquired. Specifically, suppose that we observe a pre-trained model $f^{(\ptr)}$, a target region $\mathcal{X}$, and unlabeled covariates $\tilde{\bx}_1,\ldots,\tilde{\bx}_N\in\mathcal{X}$, and that we are allowed to query labels for only $n\leq N$ of these points. We consider the setting where $n$ can be smaller and much smaller than $N$.

In this setting, we can design a sampling procedure that selects $n$ covariate points from the unlabeled sets so that their empirical distribution approximates a target sampling density $p_X(\cdot)$. The resulting retrieval algorithm is described in Algorithm~\ref{alg-sample2}.


\begin{algorithm}[H]
\begin{flushleft}
\textbf{Input:} Sampling budge $n$, target region $\calX$, and unlabeled samples $\bx^{(u)}_i\in\calX$, $i=1,\dots,N$.

\textbf{Output:} Retrieved samples $(\bx_i^{\top},y_i)$, $i=1,\dots,n$, $\mathcal{N}^{(tr)}$, and $\mathcal{N}^{(va)}$.

\textbf{Step 1}. Randomly sample $n_0$ points from $\{\bx^{(u)}_i\}_{i=1}^N$ and denote the selected index set as $\mathcal{N}_0$. Compute $\hat{\sig}^2(\bx)$ as in Step 1 of Algorithm \ref{alg-sample1} based on $\{1,\dots,n_0/2\}$. 

\textbf{Step 2}. Randomly sample $\check{\bx}_i$, $i=1,\dots,N-n_0$ points from $\calX$ by Step 2 of Algorithm \ref{alg-sample1}. 
Run logistic regression based on samples $((\bx^{(u)}_i)^{\top},0),i\in[N]\setminus\mathcal{N}_0$ and $(\check{\bx}_i^{\top},1),i=1,\dots,N-n_0$ and obtain the coefficient estimate $\hat{\bm\beta}$.
Let $\hat{r}(\bx)=\exp\{\bx^{\top}\hat{\bm\beta}\}$.

\textbf{Step 3}. Sample $n-n_0$ points from $\bx^{(u)}_i$, $i\in[N]\setminus\mathcal{N}_0$ according to weights $\hat{r}(\bx^{(u)}_i)/\sum_{i\in[N]\setminus\mathcal{N}_0}\hat{r}(\bx^{(u)}_i)$. Let $(\bx_i^{\top},y_i)$, $i=1,\dots,n$ denote the covariates sampled in Step 1 and Step 3 and their corresponding responses. Define $\mathcal{N}^{(tr)}=\{n_0+1,\dots,n\}$ and $\mathcal{N}^{(va)}=\{n_0/2+1,\dots,n_0\}$.
\end{flushleft}
\caption{Personalized nonparametric estimate given $N$ unlabeled samples from $\calX$.}
\label{alg-sample2}
\end{algorithm}

Algorithm~\ref{alg-sample2} proceeds as follows. A small number of unlabeled points is first selected uniformly to estimate the variance function and to construct a validation set. The remaining labeled samples are then chosen using an importance sampling strategy \citep{kloek1978bayesian}. Let $q_X(\bx)$ denote the distribution of the unlabeled covariates $\bx_i^{(u)}$. The estimated ratio $\hat{r}(\bx)$ serves as an approximation to $p_X(\bx)/q_X(\bx)$, enabling sampling from a distribution close to $p_X$ using only the unlabeled data. 

We now show that, under mild conditions, the proposed retrieval scheme achieves the same variance order as the optimal sampling distribution.
\begin{condition}[Conditions on density ratio]
\label{cond3}
Suppose that $\bx_i^{(u)}, i=1,\dots,N$ are i.i.d. from some distribution $q_X(\bx)$ such that $\sup_{\bx\in\calX}q_X(\bx)\leq C$ and $p^*_X(\bx)/q_X(\bx)=\exp\{\bx^{\top}\bm\beta\}$ for some $\|\bm\beta\|_2\leq C$ and $C$ is a positive constant. Moreover, the covariance matrix of $\bx_i^{(u)}$, denote by $\Sigma^{(u)}$, satisfies $\Lambda_{\min}(\Sigma^{(u)})\geq c_0>0$ for some positive constant $c_0$. 
\end{condition}
\begin{lemma}[Rate optimality of the retrieval scheme in Algorithm \ref{alg-sample2}]
\label{lem-retrieval2}
Assume Conditions \ref{cond1}, \ref{cond-noise}, and \ref{cond3}. 
Suppose that $N\geq n$, $h=n^{-c_0}$ for some constant $c_0\leq 0$, and $|\calX_1|\leq c\min\{\bar{\sig}^2/(\tilde{r}^2_nnh^d),1/\sqrt{\log n}\}$ for some positive constant $c<1$. Then there exists some positive constant $C$ such that
\[
\int_{ \calX}\E[\frac{\sig^2(\bx)+\theta_1^{(\sig)}h^{\theta_2^{(\sig)}}}{\max\{1,n_h(\bx)\}}]d\bx\leq C\frac{\bar{\sig}^2}{nh^d}.
\]
\end{lemma}
Lemma~\ref{lem-retrieval2} shows that the retrieval scheme in Algorithm~\ref{alg-sample2} achieves the same variance order as the optimal retrieval distribution $p_X^*(\bx)$ by weighted sampling from unlabeled covariates.

\section{Simulation studies}
\label{sec-simu}
In this section, we conduct multiple numerical studies to evaluate the empirical performance of the proposed method.
\subsection{Regression}
\label{sec-simu1}
In the first experiment, we set the true model to be
\[
   f^*(\bx)=\theta^*_1|x_1|+\theta^*_1|x_2+0.3|^{\theta^*_2}
\]
for $\btheta^*=(1,0.5)^{\top}$ and the target region is $\calX=[-0.5,0.5]^2$. For each retrieved $\bx_i$ from the target model, we generate its response as $y_i=f^*(\bx_i)+\eps_i$, where $\eps_i\sim N(0,1)$ independently, $i=1,\dots,n$.

\begin{figure}[H]
\includegraphics[height=5.5cm,width=6.5cm]{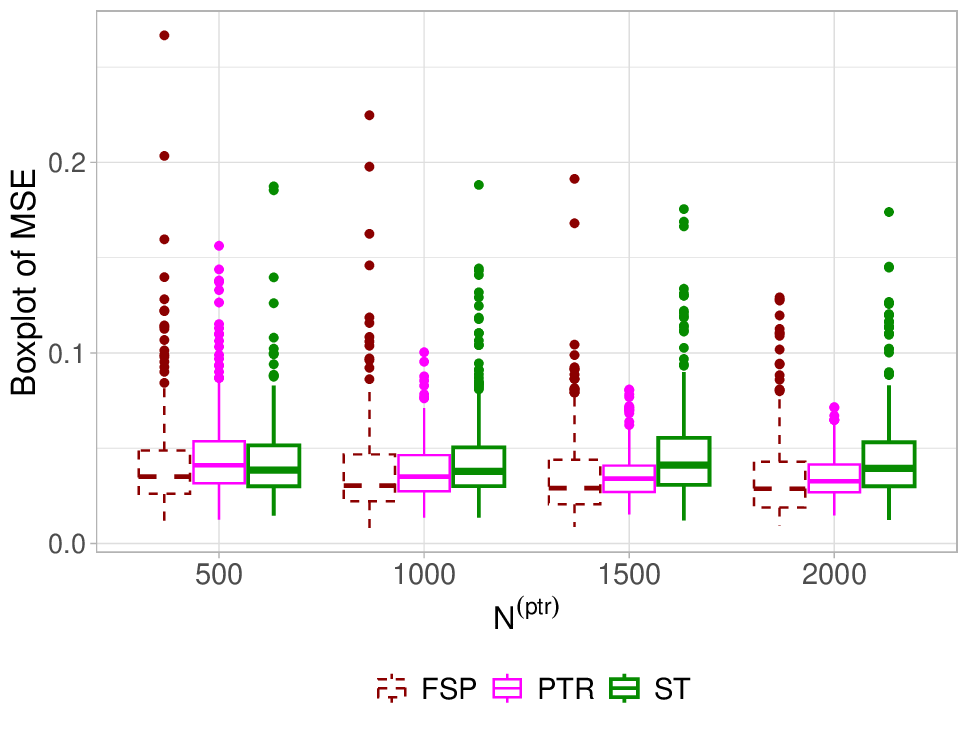}
\includegraphics[height=5.5cm,width=6.5cm]{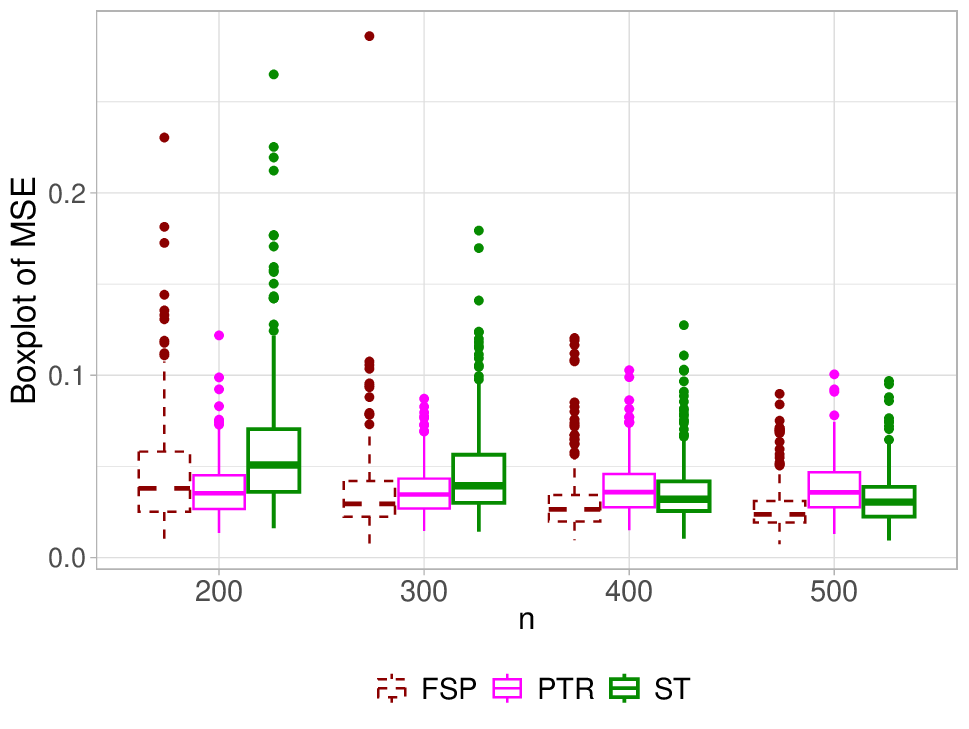}
\caption{Boxplot of the MSE for the single-task method (ST), proposed method (FSP), and the pre-trained model (PTR) with different pre-trained sample size $N^{(\ptr)}$. The sampling budget  is fixed at $n=300$ in the left plot and the pre-trained sample size if fixed at $N^{(\ptr)}=1000$ in the right plot.}
\label{fig-reg}
\end{figure}

For the pre-trained model, we sample $\bx^{(\ptr)}_i$, $i=1,\dots,N^{(\ptr)}$ uniformly randomly  from $[-1,1]^2$ and
\[
y^{(\ptr)}_i=0.8f^*(\bx^{(\ptr)}_i)+\eps^{(\ptr)}_i
\]
for $\eps^{(\ptr)}_i\sim N(0,1)$.  We estimate $f^{(\ptr)}$ via kernel regression based on $\bx^{(\ptr)}_i$ and $y_i^{(\ptr)}$, $i=1,\dots,N^{(\ptr)}$. 
For comparison, we also evaluate the performance of pre-trained model and the single-task method, where the latter one computes the regression function solely based on $n$ randomly retrieved samples from $\calX$. For the proposed method, we set $n_0=n/4$. To alleviate the efficiency loss caused by sample splitting, we do not further split the first $n_0$ samples. Instead, we use all the samples $(\bx_i^{\top},y_i)$, $i\in[n_0]$ to compute $\hat{\sig}^2(\bx)$ and also use them as the validation samples.
For evaluation, we randomly sample test covariates $\bx_i^{(t)}$ from $\calX$, $i=1,\dots,n_t$ with $n_t=500$ and compute the true conditional mean $f^*(\bx_i^{(t)})$. For an arbitrary estimator $f(\cdot)$, we report its mean estimation error
\[
   \textup{MSE}(f)=\frac{1}{n_t}\sum_{i=1}^{n_t}\{f^*(\bx_i^{(t)})-f(\bx_i^{(t)})\}^2.
\]
We consider different levels of $N^{(\ptr)}$ and sampling budget $n$.
For each setting, we repeat the above experiment independently for 300 times and report the results in Figure \ref{fig-reg}. The code is available at \url{https://github.com/saili0103/FSP}.

From the left panel of Figure \ref{fig-reg}, the MSE for both the pre-trained and proposed estimates decreases as  $N^{(\ptr)}$  increases. This is because they both leverage pre-retained samples, which are informative for the target task. The MSE of single-task method is unchanged as it does not integrate the pre-trained estimate. 
From the right panel, we see that as the target sample size $n$ grows, the MSE for the single-task and proposed estimates decreases, since both utilize the retrieved samples from the target domain. 
These results demonstrate the benefit of integrating pre-trained models and the effectiveness of our proposal in correcting the bias of the pre-trained estimate.
\subsection{Classification}
\label{sec-simu2}
\begin{figure}[H]
\includegraphics[height=5.5cm,width=6.5cm]{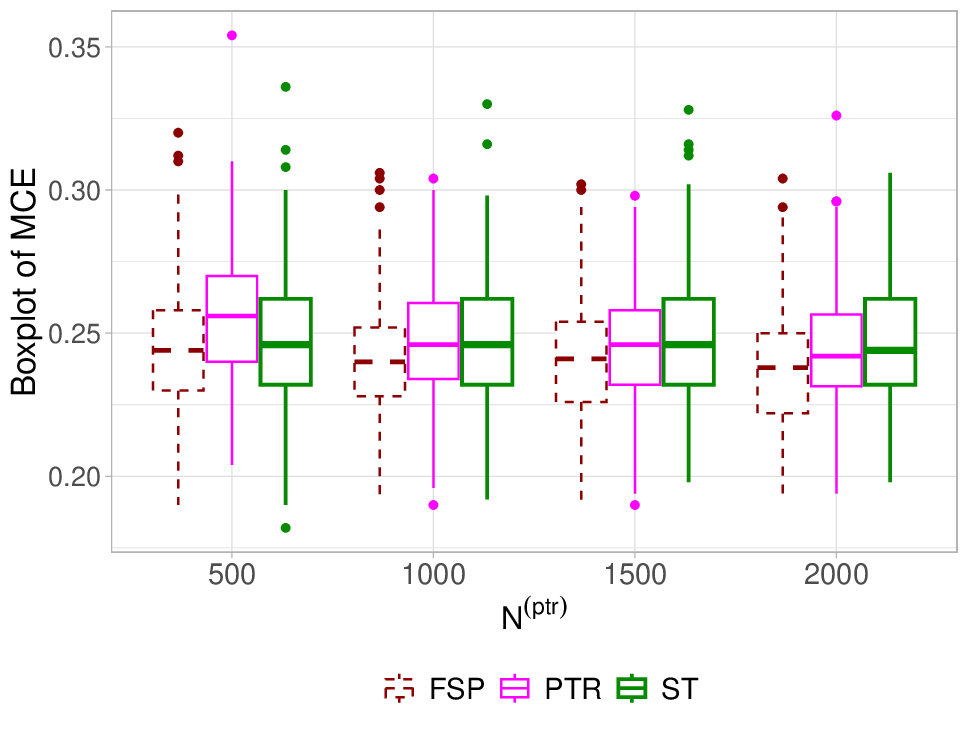}
\includegraphics[height=5.5cm,width=6.5cm]{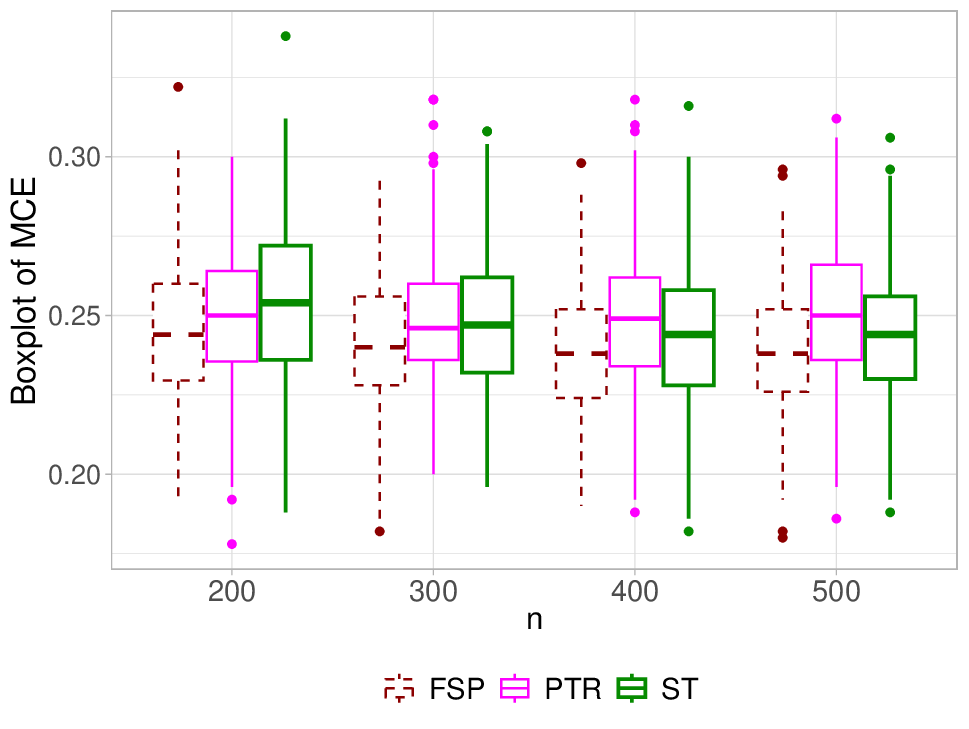}
\caption{Boxplot of the MCE for the single-task method (ST), proposed method (FSP),  and the pre-trained estimate (PTR) with different pre-trained sample size $N^{(\ptr)}$. The sampling budget $n=300$ in the left plot and the pre-trained sample size $N^{(\ptr)}=1000$ in the right plot.}
\label{fig-class}
\end{figure}

We further consider a classification task $\P(y_i=1|\bx_i)=f^*(\bx_1)$, where 
\[
f^*(\bx)=\max(\min(\theta^*_1|x_1|^{\theta^*_2}+\theta^*_1|x_2-0.3|^{\theta^*_2}-0.1,0.9),0)
\]
and $\bm\theta^*=(1, 0.6)^{\top}$.
We consider the target region $\calX=[-0.2,0.8]^2$. For the pre-trained model, we first generate $\bx^{(\ptr)}_i$ uniformly from $[-1,1]^2$ and  
\[
  \P(y^{(\ptr)}_i=1|\bx^{(\ptr)}_i)=f^*(\bx^{(\ptr)}_i)+0.1.
\]
Analogous to Section \ref{sec-simu1}, we generate test covariates $\bx_i^{(t)}$, $i=1,\dots,n_t$ randomly from $\calX$ and generate their response $y_i^{(t)}$ as Bernoulli random variables such that $\P(y_i^{(t)}=1|\bx_i^{(t)}=\bx)=f^*(\bx)$.
For an arbitrary estimator $f(\cdot)$, its mean mis-classification error is defined as
\[
   \textup{MCE}(f)=\frac{1}{n_t}\sum_{i=1}^{n_t}\left|y_i^{(t)}-\mathbbm{1}(f(\bx_i^{(t)})\ge0.5)\right|.
\]
We see from Figure \ref{fig-class} that the proposed method has smaller mean mis-classification errors than the other two methods in all the settings. The general patterns in the plots are analogous to those in Figure \ref{fig-reg} and the performance of our proposal aligns with our theory.

\subsection{Integrating noninformative pre-trained models}
\label{sec-simu3}
We evaluate the robustness of proposed method when the pre-trained model is not informative. We set the true model $f^*(\cdot)$ as in Section \ref{sec-simu1} but for the pre-trained model, we generate $f^{(\ptr)}(\bx_i)\sim N(0,1)$ independently. In this case, naively integrating pre-trained models can induce larger errors to the estimation.
\begin{table}[!htbp]
\begin{tabular}{|c|c|c|c|c|c|c|c|}
\hline
$N^{(\ptr)}$ &ST & FSP & PTR\\
\hline
500 &0.043 (0.02)&0.051 (0.02)&0.712 (0.14) \\
1000& 0.047 (0.03)&0.048 (0.02)&0.700 (0.10)\\
1500 & 0.045 (0.03)&0.046 (0.02)&0.707 (0.08)\\
2000&0.046 (0.03)&0.043 (0.02)&0.703 (0.07)\\
\hline
\end{tabular}
\begin{tabular}{|c|c|c|c|}
\hline
$n$ &ST & FSP & PTR \\
\hline
200 & 0.056 (0.03) &0.063 (0.03) & 0.703 (0.08)\\
300 & 0.046 (0.02) & 0.047 (0.02) & 0.693 (0.09)\\
400 & 0.037 (0.02) &0.039 (0.02) &0.713 (0.10)\\
500 &0.032 (0.02) &0.035 (0.01) &0.698 (0.09)\\
\hline
\end{tabular}
\caption{The mean (standard deviation) of MSE for the single-task method (ST), proposed method (FSP), and pre-trained model (PTR) based on 300 Monte Carlo simulations. The sampling budget $n=300$ in the left table and the pre-trained sample size $N^{(\ptr)}=1000$ in the right table.}
\label{tab1}
\end{table}

In Table \ref{tab1}, we report the MSE of different methods in this setting. We see that the proposed method has estimation errors comparable to the errors of single-task methods. The results show that the proposed method is robust to the adversarial pre-trained models, which aligns with our theoretical analysis.

\section{Real data study}
\label{sec-data}
In this section, we apply the proposed method to predict the housing prices in California.

The dataset contains 20640 records from the 1990 U.S. Census at the census block group level. The goal is to predict the median housing price of a block using nine features. 
This data was initially featured in \cite{pace1997sparse} and is available at \url{https://www.kaggle.com/datasets/camnugent/california-housing-prices}.
We study predicting the median housing price for near bay region, which has 2290 samples, based on five features: block longitude(lon), block latitude(lat), median age of houses in the block(age), total population of the block group(pop), and median household income in the block(inc).
 We leave out 1000 samples as test data. For the single-task method, we randomly sample 500 records from the rest 1290 samples to build the nonparametric regression model.
For personalization, we first pretend the 1290 samples are all unlabeled and run Algorithm \ref{alg-sample2} to retrieve $n=500$ samples. We use these 500 labeled samples as the retrieval data. 
 
 We consider three pre-trained models. For first one, we describe this prediction task to DeepSeek-V3.2 and ask the AI model to give a formula for estimating the median housing price in California. The detailed prompt and AI response are available at 
 \url{https://chat.deepseek.com/share/q55zlavzp48lrc9ntm}. The prediction rule given by DeepSeek-V3.2 is
 \[
 20+15\times\text{inc}-0.3\times\text{age}+0.005\times\text{pop}-1.5\times\text{lon}-0.8\times\text{lat}+0.02\times\text{lat}^2-0.1\times\text{inc}\times\text{lon}.
 \]
For the second pre-trained model, we train a random forest (RF) \cite{breiman2001random} using the samples outside the Bay area in California with sample size 18350. For the third pre-trained model, we consider LightGBM \cite{ke2017lightgbm}, a popular gradient boosting framework that uses tree based learning algorithms, and also train it based on samples outside the Bay area.

 \begin{figure}[H]
 \includegraphics[scale=0.5]{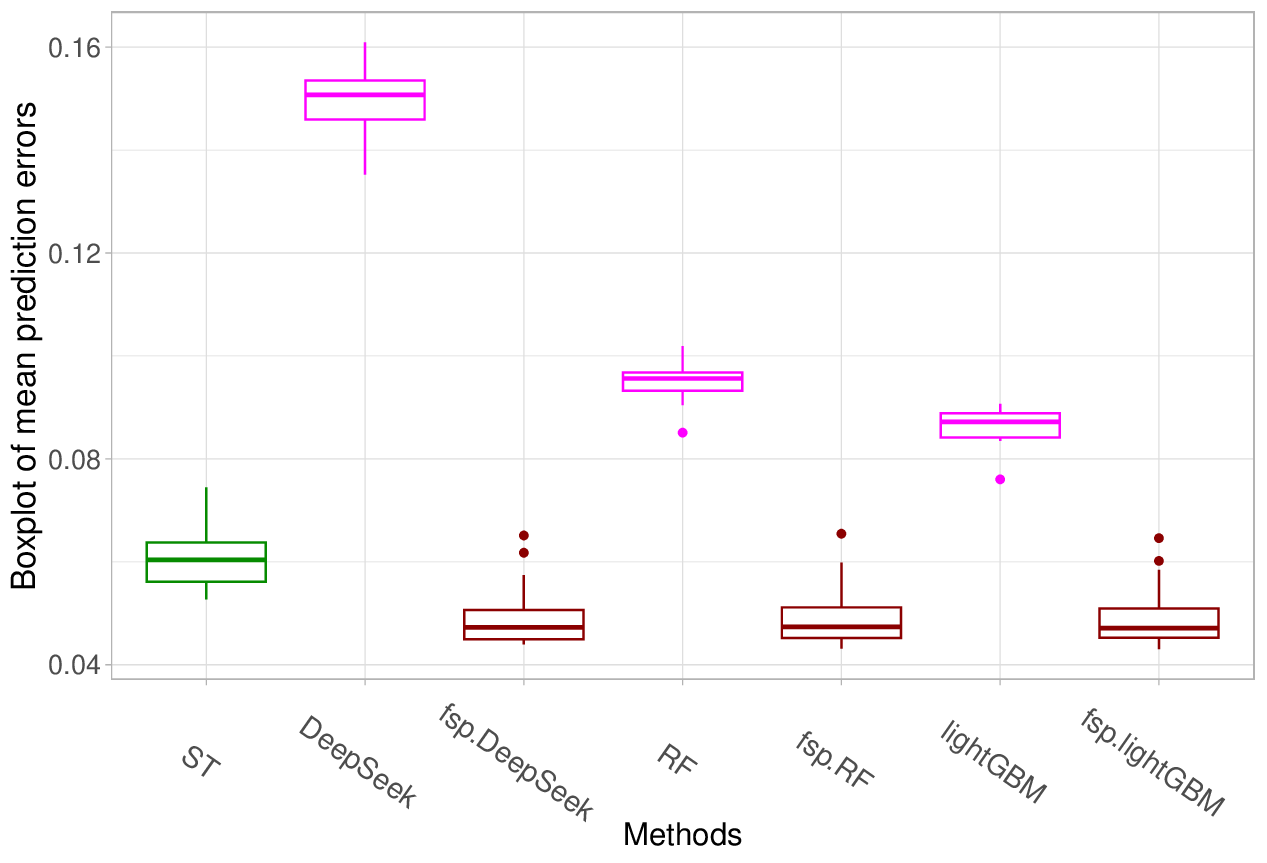}
  \caption{Boxplots of mean prediction errors for each method. Each boxplot is based on 20 random splits of test and training samples. The methods in comparison include single-task nonparametric regression (ST), pre-trained models (DeepSeek, RF, and lightGBM), and few-shot personalized methods (fsp.DeepSeek, fsp.RF, and fsp.lightGBM).}
  \label{fig-data}
 \end{figure}
The results for these experiments are presented in Figure \ref{fig-data}.  
We see that the pre-trained models all have larger prediction errors than the single-task method. It implies that the model deduced by DeepSeek is inaccurate and there exist distribution shifts in the housing price between the Bay area and other areas in California. Especially, the model given by DeepSeek, which is largely data independent, is worse than the other data-driven models. Nevertheless, after the proposed personalization method, the predictive performance of all pre-trained models improves substantially, surpassing the accuracy of the single-task methods. 
This analysis shows that the proposed personalization scheme can efficiently borrow information from other datasets and incorporate knowledge from other domains, such as geography and economics. Therefore, personalization offers a way to reduce the data collection costs required for the target domain.

\section{Discussion}
In this work, we present a statistical framework for few-shot personalization and develop a minimax rate optimal personalization method for nonparametric regression. We show that integrating large-scale pre-trained models can achieve better estimation accuracy than training from scratch and maintain robustness at the same time. The problem of personalization can also be studied for other statistical purposes. For instance, it is of interest to study personalization for estimation and prediction in parametric models, such as high-dimensional linear models and generalized linear models. 
\bibliographystyle{imsart-number} 
\bibliography{RAG.bib}
\end{document}